\documentclass[pra,reprint,twocolumn,amsmath,amssymb,superscriptaddress]{revtex4-1}
\usepackage{color}

\usepackage{graphicx}
\usepackage{epsfig}
\usepackage{bm}
\usepackage{soul}
\usepackage[normalem]{ulem}

\newcommand{\bra}[1]{\ensuremath{\left\langle{#1}\right\vert}}

\newcommand{\ket}[1]{\ensuremath{\left|{#1}\right\rangle}}

\def\bea{\begin{eqnarray}}
\def\eea{\end{eqnarray}}

\begin{document}

\title{Systematic coarse-graining of environments for \\ the non-perturbative simulation of open quantum systems}

\author{Nicola Lorenzoni}
\affiliation{Institut f\"ur Theoretische Physik und IQST, Albert-Einstein-Allee 11, Universit\"at Ulm, D-89081 Ulm, Germany}
\author{Namgee Cho}
\affiliation{Institut f\"ur Theoretische Physik und IQST, Albert-Einstein-Allee 11, Universit\"at Ulm, D-89081 Ulm, Germany}
\author{James Lim}
\affiliation{Institut f\"ur Theoretische Physik und IQST, Albert-Einstein-Allee 11, Universit\"at Ulm, D-89081 Ulm, Germany}
\author{Dario Tamascelli}
\affiliation{Institut f\"ur Theoretische Physik und IQST, Albert-Einstein-Allee 11, Universit\"at Ulm, D-89081 Ulm, Germany}
\affiliation{Dipartimento di Fisica ``Aldo Pontremoli'', Universit{\`a} degli Studi di Milano, Via Celoria 16, 20133 Milano-Italy}
\author{Susana F. Huelga}\email{susana.huelga@uni-ulm.de}
\affiliation{Institut f\"ur Theoretische Physik und IQST, Albert-Einstein-Allee 11, Universit\"at Ulm, D-89081 Ulm, Germany}
\author{Martin B. Plenio}\email{martin.plenio@uni-ulm.de}
\affiliation{Institut f\"ur Theoretische Physik und IQST, Albert-Einstein-Allee 11, Universit\"at Ulm, D-89081 Ulm, Germany}

\begin{abstract}
Conducting precise electronic-vibrational dynamics simulations of molecular systems poses significant challenges when dealing with realistic environments composed of numerous vibrational modes. Here, we introduce a technique for the construction of effective phonon spectral densities that capture accurately open system dynamics over a finite time interval of interest. When combined with existing non-perturbative simulation tools, our approach can reduce significantly the computational costs associated with many-body open system dynamics.
\end{abstract}

\maketitle

{\it Introduction.} The interaction between electronic and vibrational degrees of freedom governs various dynamical processes in molecular complexes, such as energy and charge transfer~\cite{May2011,JangPRL2004,RengerBJ2006,PriorCH+10, WomickJPCB2011,RomeroNature2017,JangRMP2018,MattioniPRX2021,SomozaarXiv2022} and chirality-induced spin selectivity~\cite{ZhangPRB2020,DuPRB2020,FranssonPRB2020,VittmannJPCL2023}. This electron-vibrational (vibronic) interaction is typically the result of a quasi-continuous low-frequency vibrational spectrum and several tens of underdamped high-frequency vibrational modes, whose phonon energies
are comparable to or larger than the thermal energy at room temperature. Electronic states interact not only with near-resonant vibrational modes~\cite{PriorCH+10,ChinNP2013,TiwariPNAS2013,PlenioAH13, ChenuSR2013,RomeroNP2014,FullerNC2014,ButkusJCP2014,LimNC2015}, but also with multiple modes over a broad frequency range with vibronic coupling strengths beyond the weak coupling regime. This is the case even for biological photosynthetic pigment-protein complexes (PPCs)~\cite{Caycedo2022} where the vibronic coupling is generally weaker than that achievable in engineered molecular systems, such as synthetic dyes and aggregates~\cite{SpanoACR2010,FarouilJPCA2018,BarclayJPCL2022}. 

Vibrational environments characterised by a high degree of structure and extending well beyond the weak coupling regime, have made it imperative to employ non-perturbative methods for simulating electronic dynamics and optical responses. Nowadays, a variety of non-perturbative tools designed to address different system types are in widespread use. The time-evolving density operator with orthogonal polynomials algorithm (TEDOPA)~\cite{PriorCH+10,TamascelliPRL2019,TamascelliMC2022} and thermofield-based chain mapping approaches~\cite{deVegaPRA2015,ChenPRE2020} allow to fully consider highly-structured vibrational environments, but they have been applied primarily to small electronic systems such as dimers. The multilayer extension of the multiconfiguration time-dependent Hartree (ML-MCTDH) method~\cite{MeyerCPL1990,WangJCP2003} tackles systems consisting of dozens of electronic states and discrete vibrational modes, typically at zero temperature as computationally expensive statistical sampling is required to account for finite temperature effects. Continuous phonon spectral densities of the spin-boson model at zero temperature have been described by a few hundred discrete modes in the weak coupling regime, but several thousand modes in the strong coupling regime to obtain numerically exact results~\cite{WangNJP2008}. For multi-site PPCs at zero temperature, the low energy parts of continuous phonon spectral densities of PPCs have been considered approximately by using several tens of discrete modes per site in ML-MCTDH simulations~\cite{SchulzeJPCB2015,SchulzeJCP2016,SchulzeCP2017,ShiblJPB2017}, but convergence to the exact results~\cite{SchroterPR2015} and its extension to full phonon environments remain to be assessed. The time-evolving matrix product operators (TEMPO) method~\cite{StrathearnNC2018} and transfer-tensor-TEDOPA~\cite{RosenbachCH+2015} yield efficiency improvements when applied to vibrational environments with correlation times shorter than or comparable to the timescale of system dynamics. The hierarchical equation of motion (HEOM)~\cite{TanimuraJPSJ1989} approach has been mainly employed to consider a broad vibronic coupling spectrum consisting of a few peaks, as its computational costs rapidly increase with the number of exponentials in the bath correlation function (BCF). The highly-structured vibrational environments of molecular systems have been severely coarse-grained in HEOM simulations~\cite{IshizakiPNAS2009,StruempferJCP2009,KreisbeckJPCL2012,KreisbeckJCTC2014,BlauPNAS2018} based on two criteria~\cite{BlauPNAS2018}: the conservation of the total vibronic coupling strength, quantified by reorganization energy, and the agreement of monomer absorption spectra computed based on actual and coarse-grained phonon spectral densities, which can be readily computed in a non-perturbative manner. However, the validity of the coarse-graining scheme has never been rigorously tested for multi-chromophoric systems using non-perturbative methods.

In this work, we leverage the observation that the complexity of the BCF, the key determinant of open-system dynamics, increases with time in the presence of highly-structured spectral densities. We develop a systematic and reliable method for constructing effective environments that describe system dynamics accurately within a chosen finite time interval of interest. When combined with the dissipation-assisted matrix product factorization (DAMPF)~\cite{SomozaPRL2019}, a non-perturbative method for simulating open-system dynamics, our approach is capable of capturing the influence of environmental fluctuations on electronic system dynamics, while considerably reducing the computational costs. We demonstrate the efficiency of the method by presenting the absorption spectra of the Fenna-Matthews-Olson (FMO) photosynthetic complex at finite temperature, computed for the first time in an accurate and non-perturbative manner based on experimentally estimated phonon spectral density~\cite{RatsepJL2007}. Furthermore, we show that conventional coarse-graining schemes can fail to reproduce the absorption spectra based on actual vibrational environments, even qualitatively. This underscores the importance of systematically constructing effective environments in open-system simulations. Importantly, our method extends beyond the simulation of absorption, enabling one to reduce the computational costs of non-perturbation simulations of general open-system dynamics.
\vspace{0.2cm}

{\it Model.}
Considering a multi-chromophoric system consisting of $N$ pigments, we express the electronic Hamiltonian within the single excitation subspace as $H_e=\sum_{i=1}^{N}\epsilon_i\ket{\epsilon_i}\bra{\epsilon_i}+\sum_{i\neq j}^{N}V_{ij}\ket{\epsilon_i}\bra{\epsilon_j}$, where $\ket{\epsilon_i}$ denotes a local electronic excitation of pigment $i$ with site energy $\epsilon_i$ and $V_{ij}$ an intra-pigment electronic coupling. Here we assume that higher-energy excited states do not contribute to the low-energy part of linear optical spectra, which is a reliable description of PPCs consisting of bacteriochlorophylls such as the FMO complex~\cite{RengerBJ2006}. Considering an ensemble of PPCs, the site energies $\epsilon_i$ of pigments can vary due to non-identical local environments, resulting in static disorder. This can be addressed by sampling the site energies $\epsilon_i$ from independent Gaussian distributions with mean values $\langle\epsilon_i\rangle$ and a standard deviation of $\sigma$, here taken as $\sigma = 80\,{\rm cm}^{-1}$, a typical value for PPCs~\cite{RengerBJ2006, Caycedo2022}.

\begin{figure}
\includegraphics[width=0.48\textwidth]{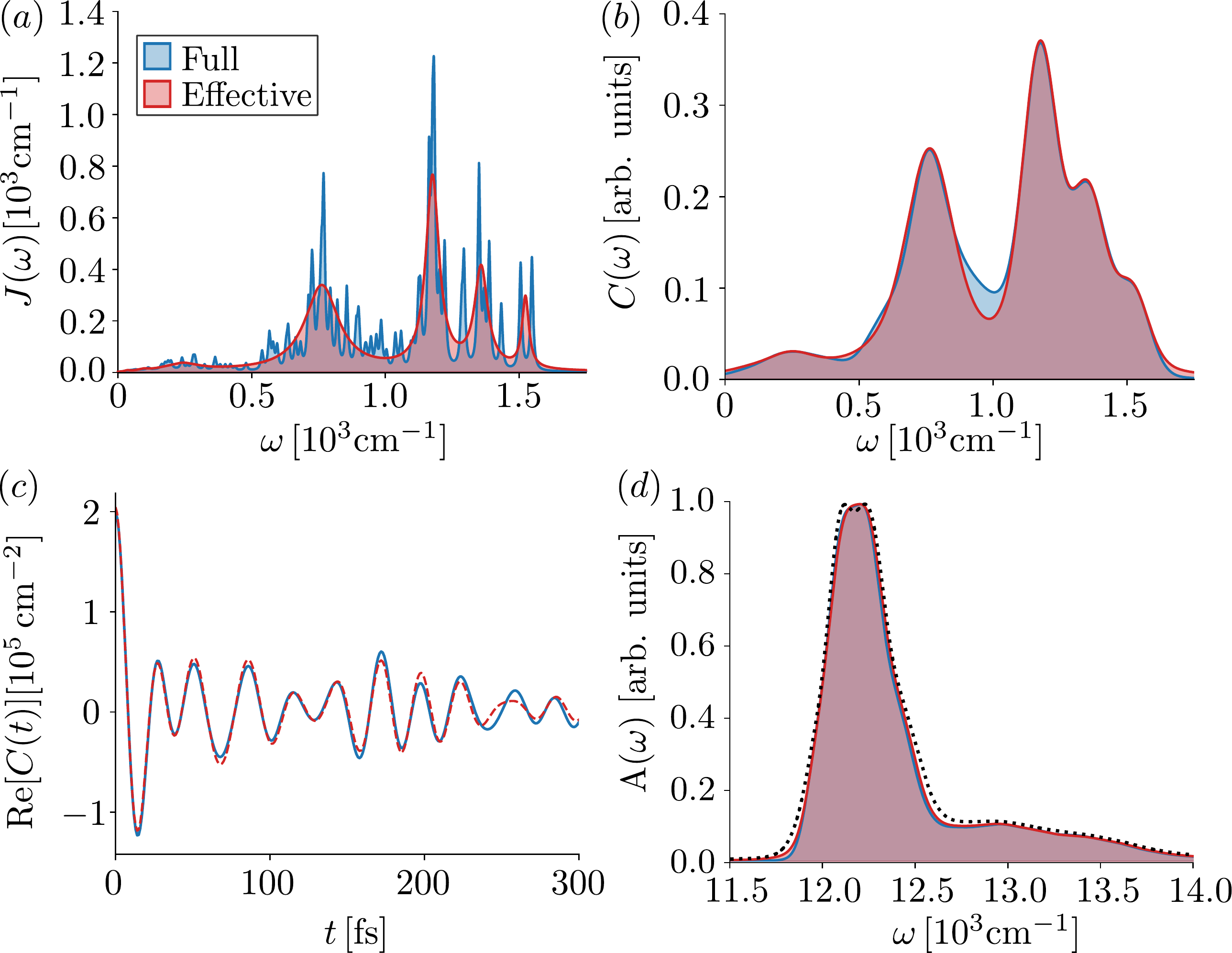}
\caption{(a) Experimentally estimated phonon spectral density of the FMO complex~\cite{RatsepJL2007} (blue) and effective spectral density consisting of six Lorentzian peaks (red). (b) FT spectrum of the Gaussian-filtered BCF of the full FMO spectral density at $77\,{\rm K}$ (blue) and that of the effective spectral density (red). (c) BCFs of the full FMO (blue) and effective (red) spectral densities without the Gaussian filter. (d) Absorption spectra of a seven-site FMO model at $77\,{\rm K}$, computed by DAMPF based on the full FMO (blue) and effective (red) spectral densities where mean site energies and Gaussian broadening were considered (see the main text). The FMO absorption spectra computed based on randomly generated site energies with the effective spectral density are shown in a black dashed line.}
\label{Fig1}
\end{figure}

The vibronic interaction is modeled by $H_{e-v}=\sum_{i=1}^{N}\ket{\epsilon_i}\bra{\epsilon_i}\sum_{k}\omega_k\sqrt{s_k}(b_{i,k}+b_{i,k}^{\dagger})$ where $b_{i,k}$ and $b_{i,k}^{\dagger}$ denote the annihilation and creation operators, respectively, of a vibrational mode with frequency $\omega_k$, locally coupled to pigment $i$ with a coupling strength quantified by the Huang-Rhys factor $s_k$. $H_{e-v}$ is fully characterized by a phonon spectral density $J(\omega)=\sum_{k}\omega_{k}^{2}s_k\delta(\omega-\omega_k)$, and the total vibronic coupling strength is quantified by the reorganization energy $\int_{0}^{\infty}d\omega J(\omega)/\omega$. In this work, we consider the experimentally estimated phonon spectral density of the FMO complex~\cite{RatsepJL2007} shown in blue in Fig.~\ref{Fig1}(a), consisting of a quasi-continuous protein spectrum, modeled via the Adolphs-Renger (AR) spectral density $J_{\rm AR}(\omega)$, and 62 intra-pigment vibrational modes, each modeled by a narrow Lorentzian spectral density (see the SM). Given the vibrational Hamiltonian $H_v=\sum_{i=1}^{N}\sum_{k}\omega_k b_{i,k}^{\dagger}b_{i,k}$, the total Hamiltonian of the PPC is given by $H=H_e+H_v+H_{e-v}$.

{\it BCF-based coarse-graining.} Absorption spectra are determined by the time evolution of optical coherences between electronic ground and excited states that are created by external light. The broadening of absorption line shapes is induced by a finite lifetime $\tau$ of the optical coherences, determined by vibronic couplings $H_{e-v}$ and static disorder, which is approximately $\tau\approx 300\,{\rm fs}$ for the FMO complex and dimeric systems at $T=77\,{\rm K}$ that we consider in this work.

When an open system couples linearly to a harmonic environment, initially in thermal equilibrium at temperature $T$, the influence of the environment on the reduced system dynamics is fully determined by the BCF $C(t)=\int_{0}^{\infty}d\omega J(\omega)(\coth(\omega/2k_B T)\cos(\omega t)-i\sin(\omega t))$~\cite{FeynmanANNPHYS1963}. More specifically, the reduced system density matrix $\rho_{s}(\tau)$ at a finite time $\tau$ is solely determined by $C(t)$ for $0\le t\le \tau$, without any influence from $C(t)$ for $t>\tau$. Therefore, the physical quantities determined by system dynamics over a finite time $\tau$ can be simulated by using an effective spectral density instead of actual one, provided that their BCFs are well matched for $0\le t\le \tau$. This implies that for accurate simulations of the FMO absorption spectra, one needs to construct an effective phonon spectral density quantitatively describing the BCF of the full FMO spectral density up to $\tau\approx 300\,{\rm fs}$.

\begin{figure}
\includegraphics[width=0.45\textwidth]{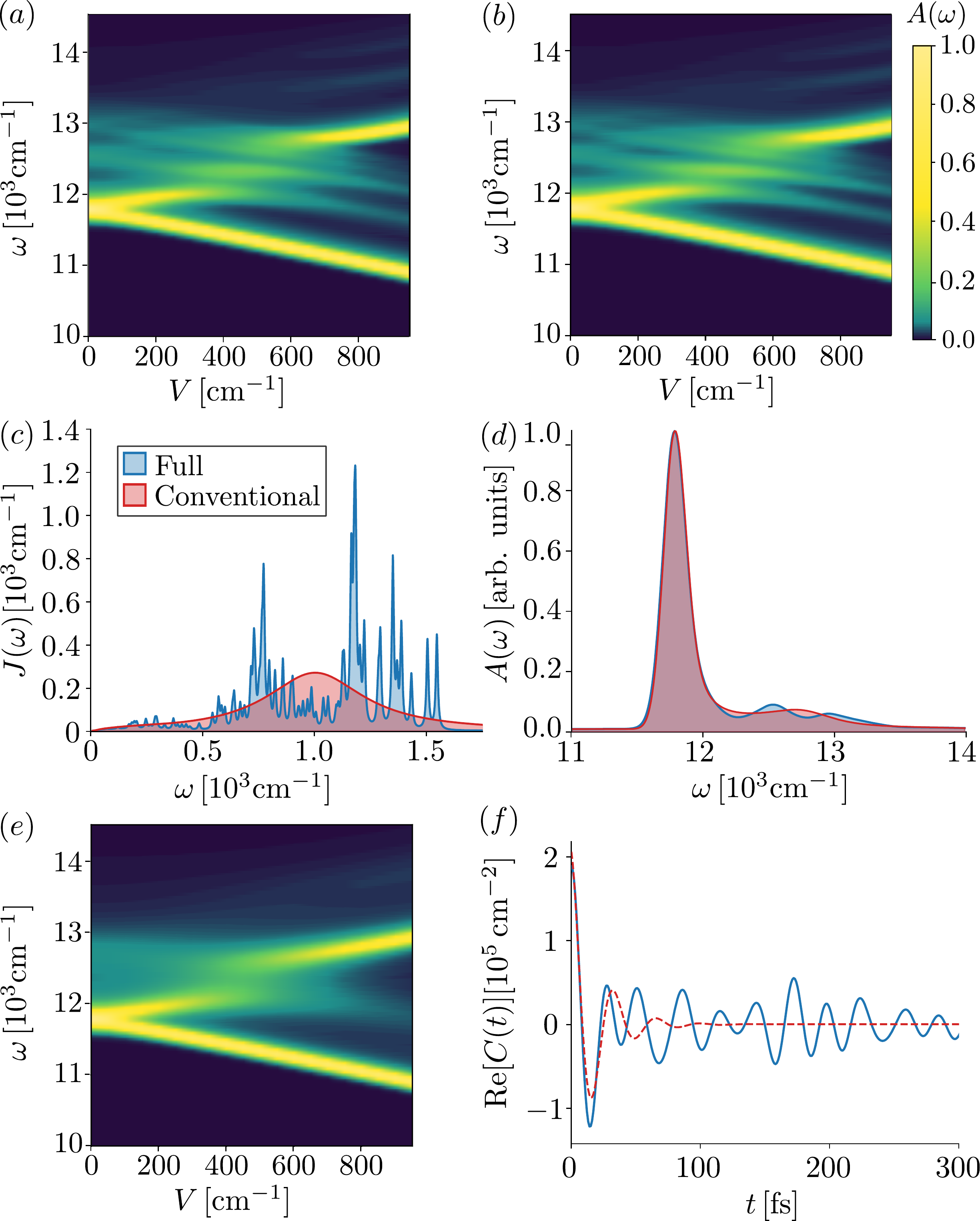}
\caption{(a,b) Dimer absorption spectra at $77\,{\rm K}$ as a function of electronic coupling $V$, computed using HEOM with (a) the full FMO and (b) effective spectral densities shown in Fig.~\ref{Fig1}(a). (c) The full FMO spectral density (blue) and conventional coarse-grained spectral density (red). (d) Monomer absorption spectra at $77\,{\rm K}$ computed based on the full (blue) and conventional coarse-grained (red) spectral densities. (e) Dimer absorption spectra at $77\,{\rm K}$, computed by HEOM with the conventional coarse-grained model. (f) BCFs of the full FMO (blue) and conventional coarse-grained (red) spectral densities at $77\,{\rm K}$.}
\label{Fig2}
\end{figure}

In Fig.~\ref{Fig1}(b), the Fourier transformed (FT) spectrum of $C(t)$ of the full FMO spectral density at $T=77\,{\rm K}$ over $0\le t\le 300\,{\rm fs}$ is shown in blue. Since a sharp cut-off of $C(t)$ at $t=300\,{\rm fs}$ leads to ringing artifacts, we applied a Gaussian filter before the FT, so that the Gaussian filtered BCF $C(t)e^{-t^2/2 \tilde{\sigma}^{2}}$ with a standard deviation $\tilde{\sigma} = 100\,{\rm fs}$ becomes negligible at $t>3\tilde{\sigma} = 300\,{\rm fs}$. The Gaussian filter also makes the FT spectrum dominated by $C(t)$ at early times, which has more impact on system dynamics than $C(t)$ at later times. Noticeably, the FT spectrum is much less-structured than the full FMO spectral density, as the former can be well reproduced by a sum of the Gaussian-filtered BCFs of only six Lorentzian spectral densities, as shown in red in Fig.~\ref{Fig1}(b), with one representing the contribution of the broad AR spectral density and five describing that of the 62 intra-pigment modes (see the SM). Here the parameters of the effective spectral density, shown in red in Fig.~\ref{Fig1}(a), were determined in such a way that the total reorganization energy and Huang-Rhys factor of the full FMO spectral density are conserved. Fig.~\ref{Fig1}(c) shows the BCFs of the effective and full FMO spectral densities, are well matched up to $t\approx 300\,{\rm fs}$. We note that a similar degree of reduction in the complexity of phonon spectral densities can be found for other systems (see the SM).

To demonstrate that the effective spectral density can describe the FMO absorption spectra accurately, we employ DAMPF~\cite{SomozaPRL2019,SomozaarXiv2022} with electronic Hamiltonian $H_e$ of the FMO complex estimated in Ref.~\cite{RengerBJ2006}. Fig.~\ref{Fig1}(d) shows the absorption spectra of the seven-site FMO model computed by DAMPF based on the full FMO and effective spectral densities, respectively, shown in blue and red, which are quantitatively well matched. Here, the static disorder was treated approximately by computing the optical coherence dynamics using the mean site energies $\langle\epsilon_i \rangle$ of the FMO complex and multiplying it by a Gaussian broadening $e^{-t^2/2\tilde{\sigma}^2}$, so that the coherence decays within $3\tilde{\sigma}=300\,{\rm fs}$. The use of the effective environmental spectral density described above reduces the computational time (memory cost) from $40$ min to $3$ min (from $400\,{\rm MB}$ to $4\,{\rm MB}$) when compared to a simulation with the full FMO environment (all simulations were executed using 15 cores in an Intel Xeon 6252 Gold CPU). Since an accurate description of static disorder requires the repetitions of optical coherence simulations with randomly generated site energies, typically requiring $\sim 10^{3}$ samples, the reduction in computational costs enables one to efficiently and accurately take into account the ensemble dephasing induced by static disorder, as shown in a black dashed line in Fig.~\ref{Fig1}(d), where the absorption spectra were computed using the effective spectral density and randomly generated site energies $\epsilon_i$. The reduction in computational costs will become even more relevant in molecular parameter estimation, where the mean site energies $\langle\epsilon_i\rangle$ and electronic couplings $V_{ij}$ are optimized until experimental absorption spectra are quantitatively reproduced in simulations~\cite{Caycedo2022}, thus requiring a much larger number of non-perturbative computations. We note that the computational advantage can become more pronounced for larger vibronic systems with stronger system-environment correlations (see the SM).

So far we have demonstrated how the effective environment can reduce the computational costs of DAMPF and still describe the full environmental effects. Here we discuss how the effective environment approach can impact on the computational costs of other existing methods. In conventional HEOM simulations~\cite{KramerJCC2018}, the information about the correlations between system and environments is encoded in a hierarchical structure of auxiliary operators whose dimensions are identical to that of a reduced system density matrix. For a typical PPC model consisting of $N$ pigments coupled to local vibrational environments modeled by $M$ Lorentzian spectral densities, the total number of auxiliary operators up to the $L$-th hierarchical layer is given by $(2NM+L)!/(2NM)!/L!$ when the Matsubara terms are not considered~\cite{LimPRL2019}. For a dimer ($N=2$), absorption simulation costs are reduced by almost 5 orders of magnitude, from $\sim 0.27\,{\rm TB}$ to $\sim 3.8\,{\rm MB}$, as the number of Lorentzians is decreased from $M=62$ to $M=6$, when a typical hierarchical depth $L=5$ is considered. For multi-chromophoric systems consisting of $N=10, 20, 30$ pigments, when our effective spectral density with $M=6$ is considered with $L=5$, the HEOM simulation costs of absorption simulations are approximately $\sim 0.038, 2.3, 25\,{\rm TB}$, respectively. This may enable HEOM simulations of functionally relevant PPC units, such as the FMO complex ($N=7$) from green sulfur bacteria~\cite{MilderPR2010}, the PC645 complex ($N=8$) from marine algae~\cite{BlauPNAS2018,ColliniNature2010}, and LH2 ($N=27$) from purple bacteria~\cite{ScholesJPCB1999,LawMMB2004}.

To demonstrate that the accuracy of the effective environment approach is insensitive to system parameters, dimer absorption spectra at $T=77\,{\rm K}$ computed by HEOM based on the full FMO and effective spectral densities are shown in Fig.~\ref{Fig2}(a) and (b), respectively, as a function of the electronic coupling $V=V_{12}$. Here the site energies $\epsilon_i$ were randomly generated with identical mean values $\langle\epsilon_1-\epsilon_2\rangle=0$, and the transition dipole moments of the two pigments were chosen orthogonal (see the SM). It is notable that the dimer absorption spectra computed with the effective environment are quantitatively well matched to the full environment model results, over the considered range of the electronic coupling strength $V$. This shows that accurate absorption spectra can be obtained by using the effective environment constructed based on BCF, independently of the parameters of system Hamiltonian.

{\it Conventional coarse-graining.} Contrary to the BCF-based effective environment, coarse-grained environments in previous HEOM studies~\cite{KreisbeckJPCL2012,KreisbeckJCTC2014,BlauPNAS2018} have been constructed using different criteria. In Fig.~\ref{Fig2}(c), a typical coarse-grained environment is shown in red, where the 62 Lorentzian peaks are replaced by a single, broader one. Here the parameters of the broad Lorentzian were determined in such a way that the total reorganization of the 62 intra-pigment modes is conserved, and the absorption spectra of monomer ($N=1$) at $77\,{\rm K}$ computed with the full and coarse-grained environments are well matched, as shown in Fig.~\ref{Fig2}(d). It is notable that the dimer absorption spectra computed by HEOM based on the coarse-grained enviroment, shown in Fig.~\ref{Fig2}(e), are even qualitatively different from the full environment model results shown in Fig.~\ref{Fig2}(a). This is due to the significant deviations between the BCFs of the coarse-grained and of the full FMO spectral density, as shown in Fig.~\ref{Fig2}(f), thus demonstrating that the conventional criteria for constructing coarse-grained environments are not sufficient to simulate open-system dynamics in a reliable way.

\begin{figure}
\includegraphics[width=0.48\textwidth]{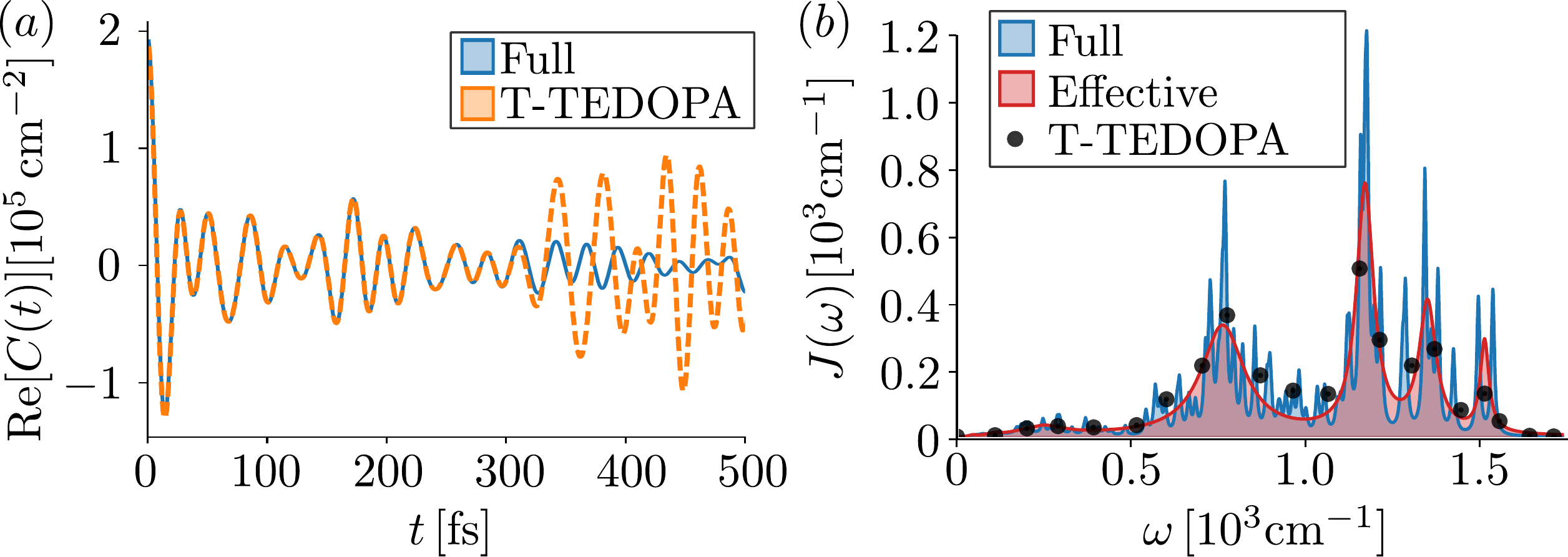}
\caption{(a) BCF of the full FMO spectral density at $77\,{\rm K}$ (blue) and that of a truncated T-TEDOPA chain (orange). (b) Discrete phonon spectral density of the truncated T-TEDOPA chain (black dots), and the full FMO (blue) and effective (red) spectral densities shown in Fig.~\ref{Fig1}(a).}
\label{Fig4}
\end{figure}

{\it TEDOPA-based coarse-graining.} So far we have demonstrated how an effective environment can be constructed for non-perturbative simulation tools where continuous spectral densities are considered. Here we discuss how an effective discrete environment can be systematically constructed for other non-perturbative techniques based on discrete modes.

In T-TEDOPA~\cite{TamascelliPRL2019}, a vibrational environment is mapped into a semi-infinite one-dimensional chain of quantum harmonic oscillators, where the electronic states couple only to the first site of the chain and every oscillator interacts only with its nearest neighbors~\cite{PriorCH+10,ChinRH+10}.
In a crucial step towards the simulation of spectral densities at finite temperature, T-TEDOPA implements a transformed spectral density determined such that all the environmental oscillators in the chain are initialized in their vacuum states and the system dynamics is provably identical to that of the original finite temperature problem~\cite{TamascelliPRL2019}. As a consequence, the first oscillator directly coupled to the electronic states is populated at early times and then the population is transferred through the semi-infinite chain. This allows a truncation of the chain in such a way that the BCF is well described within a finite time scale until the population reaches the truncated site, thus ensuring arbitrarily small error in the system observables such as spectra~\cite{MascherpaSH+17}. The finite number of oscillators of the truncated chain results in a discrete phonon spectral density~\cite{TamascelliMC2022}. Fig.~\ref{Fig4}(a) shows the BCF of the truncated T-TEDOPA chain consisting of 51 oscillators, which reproduces the BCF of the full FMO spectral density up to $300\,{\rm fs}$ (as the number of oscillators increases, the BCF can be reproduced for a longer time scale). The corresponding discrete phonon spectral density, shown in black dots in Fig.~\ref{Fig4}(b), is qualitatively similar to the effective phonon spectral density constructed based on the previously described FT-based approach with a Gaussian filter. In addition, the discrete spectral density may reduce the computational costs of other non-perturbative tools based on discrete modes. In previous ML-MCTDH simulations of PPCs~\cite{SchulzeJPCB2015,SchulzeJCP2016,SchulzeCP2017}, continuous spectral densities in the low-frequency regime were modelled by discrete modes with frequencies equally spaced within an interval of $4\,{\rm cm}^{-1}$. When applied to full phonon spectral densities with frequencies up to $\sim 2000\,{\rm cm}^{-1}$, this results in several hundreds of discrete modes per site. The number of discrete modes can be dramatically reduced when the T-TEDOPA chain is systematically truncated based on the BCF and time scale of interest, as in the case of absorption simulations requiring only several tens of discrete modes. This is in line with the numerical observations that the chain mapping outperforms other discretization schemes~\cite{deVegaPRB2015}.

\vspace{0.2cm}
{\it Conclusions.} We have developed a systematic method to construct effective vibrational environments that describe efficiently electronic dynamics in the presence of highly-structured vibrational environments within a finite time interval. When integrated with existing non-perturbative simulation tools, our approach yields a substantial reduction in computational resources, particularly when the environmental correlation time exceeds the time scale of the targeted system dynamics. With increasing simulation time scales, for example, from the sub-picosecond time scale of linear optical responses to the picosecond time scale of energy/charge transfer dynamics and nonlinear optical responses, the FT spectrum of the BCF of highly-structured vibrational environments grows increasingly complex, revealing the multimodal nature of these environments. Nonetheless, our effective environment construction requires fewer computational resources than modeling the actual environment (see the SM). The effective environment can be applied in general non-perturbative simulations of multi-partite systems coupled to harmonic environments, including polaritonic systems where emitter-photon coupling spectra are highly structured~\cite{delPinoPRL2019,
MedinaPRL2021} (see the SM). This may also support studies concerning the feasibility and resilience of proposed molecular information processors~\cite{WasielewskiNRC2020} and offer insights into the primary environmental structure responsible for open-system dynamics and its impact on experimentally observable quantities.

{\it Acknowledgements.} We thank Thomas Renger and Ivan Medina for helpful discussions and for providing us with the simulation parameters of the FMO complex and polaritonic systems. This work was support by the ERC Synergy grant HyperQ (Grant No. 856432), the BMBF
project PhoQuant (grant no. 13N16110), the Quantera project ExtraQt and the state of Baden-W{\"u}rttemberg through bwHPC and the German Research Foundation 
(DFG) through grant no INST 40/575-1 FUGG (JUSTUS 2 cluster).

\end{document}


\begin{widetext}

\newpage

\begin{center}
{\large\bf Supplemental Material: Systematic coarse-graining of environments for \\ the non-perturbative simulation of open quantum systems}
\end{center}

\section{Phonon spectral density of the FMO complex}

The experimentally estimated phonon spectral density of the FMO complex consists of a quasi-continuous spectrum of low-frequency protein modes and 62 high-frequency intra-pigment vibrational modes~\cite{SI_RatsepJL2007}. The low-frequency spectrum is modelled by the Adolphs-Renger (AR) spectral density
\begin{equation}
	J_{\rm AR}(\omega)=\frac{S}{s_1+s_2}\sum_{i=1}^{2}\frac{s_i}{7! 2 \omega_i^4}\omega^5 e^{-(\omega/\omega_i)^{1/2}},
	\label{eq:B777_SD}
\end{equation}
where $S=0.29$, $s_1=0.8$, $s_2=0.5$, $\omega_1=0.069\,{\rm meV}$ and $\omega_2=0.24\,{\rm meV}$. The intra-pigment vibrational modes are modelled by a sum of 62 Lorentzian functions
\begin{equation}
    J_{h}(\omega) = \sum_{k=1}^{62} \frac{4 \omega_k s_k \gamma_k (\omega_{k}^2+\gamma_{k}^2)\omega}{\pi((\omega+\omega_k)^{2}+\gamma_{k}^2)((\omega-\omega_k)^{2}+\gamma_{k}^2)},
\end{equation}
where the $k$-th Lorentzian peak is centered at vibrational frequency $\omega_k$ with a height being proportional to the Huang-Rhys factor $s_k$. The experimentally estimated values of the vibrational frequencies and Huang-Rhys factors~\cite{SI_RatsepJL2007} are summarized in Table~\ref{Table_intrapigment}. The width of the Lorentzian functions, determining the vibrational damping rate of the intra-pigment modes, is taken to be $\gamma_k = 1\,{\rm ps}^{-1} \approx 5\,{\rm cm}^{-1}$. The total phonon spectral density of the FMO complex is given by $J(\omega)=J_{\rm AR}(\omega)+J_{h}(\omega)$.

\begin{table}[b]
\caption{Vibrational frequencies $\omega_k$ and Huang-Rhys factors $s_k$ of the 62 intra-pigment modes of the FMO complex~\cite{SI_RatsepJL2007}.}
\label{Table_intrapigment}
\begin{ruledtabular}
\begin{tabular}{llllllllllllll}
$k$ & 1 & 2 & 3 & 4 & 5 & 6 & 7 & 8 & 9 & 10\\
\hline
$\omega_k\,[{\rm cm}^{-1}]$ & 46 & 68 & 117 & 167 & 180 & 191 & 202 & 243 & 263 & 284 \\
$s_k$ & 0.011 & 0.011 & 0.009 & 0.009 & 0.010 & 0.011 & 0.011 & 0.012 & 0.003 & 0.008 \\
\hline\hline
$k$ & 11 & 12 & 13 & 14 & 15 & 16 & 17 & 18 & 19 & 20 \\
\hline
$\omega_k\,[{\rm cm}^{-1}]$ & 291 & 327 & 366 & 385 & 404 & 423 & 440 & 481 & 541 & 568 \\
$s_k$ & 0.008 & 0.003 & 0.006 & 0.002 & 0.002 & 0.002 & 0.001 & 0.002 & 0.004 & 0.007\\
\hline\hline
$k$ & 21 & 22 & 23 & 24 & 25 & 26 & 27 & 28 & 29 & 30 \\
\hline
$\omega_k\,[{\rm cm}^{-1}]$ & 582 & 597 & 630 & 638 & 665 & 684 & 713 & 726 & 731 & 750 \\
$s_k$ & 0.004 & 0.004 & 0.003 & 0.006 & 0.004 & 0.003 & 0.007 & 0.010 & 0.005 & 0.004 \\
\hline\hline
$k$ & 31 & 32 & 33 & 34 & 35 & 36 & 37 & 38 & 39 & 40\\
\hline
$\omega_k\,[{\rm cm}^{-1}]$ & 761 & 770 & 795 & 821 & 856 & 891 & 900 & 924 & 929 & 946 \\
$s_k$ & 0.009 & 0.018 & 0.007 & 0.006 & 0.007 & 0.003 & 0.004 & 0.001 & 0.001 & 0.002\\
\hline\hline
$k$ & 41 & 42 & 43 & 44 & 45 & 46 & 47 & 48 & 49 & 50 \\
\hline
$\omega_k\,[{\rm cm}^{-1}]$ & 966 & 984 & 1004 & 1037 & 1058 & 1094 & 1104 & 1123 & 1130 & 1162 \\
$s_k$ & 0.002 & 0.003 & 0.001 & 0.002 & 0.002 & 0.001 & 0.001 & 0.003 & 0.003 & 0.009 \\
\hline\hline
$k$ & 51 & 52 & 53 & 54 & 55 & 56 & 57 & 58 & 59 & 60 \\
\hline
$\omega_k\,[{\rm cm}^{-1}]$ & 1175 & 1181 & 1201 & 1220 & 1283 & 1292 & 1348 & 1367 & 1386 & 1431 \\
$s_k$ & 0.007 & 0.010 & 0.003 & 0.005 & 0.002 & 0.004 & 0.007 & 0.002 & 0.004 & 0.002 \\
\hline\hline
$k$ & 61 & 62 \\
\hline
$\omega_k\,[{\rm cm}^{-1}]$ & 1503 & 1545 \\
$s_k$ & 0.003 & 0.003 \\
\end{tabular}
\end{ruledtabular}
\end{table}

In the main manuscript, an effective phonon spectral density constructed based on the bath correlation function of the full FMO spectral density $J_{\rm AR}(\omega)+J_{h}(\omega)$ is introduced, which can reproduce numerically exact absorption spectra of dimers computed based on the full FMO phonon spectral density. In HEOM simulations, the effective phonon spectral density was modelled by the sum of the AR spectral density and five Lorentzian functions
\begin{equation}
    J_{\rm effective}(\omega) = J_{\rm AR}(\omega)+\sum_{k=1}^{5} \frac{4 \Omega_k S_k \Gamma_k (\Omega_{k}^2+\Gamma_{k}^2)\omega}{\pi((\omega+\Omega_k)^{2}+\Gamma_{k}^2)((\omega-\Omega_k)^{2}+\Gamma_{k}^2)},
\end{equation}
where the parameters $\{\Omega_k,S_k,\Gamma_k\}$ of the Lorentzian peaks are summarized in Table~\ref{Table_L5}. In DAMPF simulations~\cite{SI_SomozaPRL2019}, the 62 intra-pigment modes were effectively described by five pseudo-modes~\cite{SI_tama18,SI_mascherpa20} with parameters in Table~\ref{Table_L5} and the AR spectral density was modelled by an additional pseudo-mode with $\Omega_6 = 160\,{\rm cm}^{-1}$, $S_6 = 0.164$ and $\Gamma_6=133\,{\rm cm}^{-1}$. The temperature of the Markovian baths coupled to the pseudo-modes was taken to be $T=77\,{\rm K}$. We note that a single pseudo-mode is sufficient to describe the influence of the AR spectral density on absorption spectra, as the simulated results do not show any appreciable changes even if the AR spectral density is modelled by 3, 5, or 15 pseudo-modes for a more accurate description (not shown here).

\begin{table}[t]
\caption{The parameters $\{\Omega_k,S_k,\Gamma_k\}$ of the five Lorentzians considered in the effective phonon spectral density $J_{\rm effective}(\omega)$.}
\label{Table_L5}
\begin{ruledtabular}
\begin{tabular}{lllllllll}
$k$ & 1 & 2 & 3 & 4 & 5 \\
\hline
$\Omega_k\,[{\rm cm}^{-1}]$ & 247 & 763 & 1175 & 1356 & 1521 \\
$S_k$  & 0.056 & 0.133 & 0.049 & 0.019 & 0.006 \\
$\Gamma_k\,[{\rm cm}^{-1}]$  & 53 & 76 & 29 & 29 & 15 \\
\end{tabular}
\end{ruledtabular}
\end{table}

In the main manuscript, a conventional coarse-grained spectral density is introduced, consisting of the AR spectral density and a single broad Lorentzian peak
\begin{equation}
    J_{\rm conventional}(\omega) = J_{\rm AR}(\omega) + \frac{4 \tilde{\Omega} \tilde{S} \tilde{\Gamma} (\tilde{\Omega}^2+\tilde{\Gamma}^2)\omega}{\pi((\omega+\tilde{\Omega})^{2}+\tilde{\Gamma}^2)((\omega-\tilde{\Omega})^{2}+\tilde{\Gamma}^2)},
\end{equation}
where $\tilde{\Omega}=1000\,{\rm cm}^{-1}$, $\tilde{S}=0.2093$, and $\tilde{\Gamma}=(20\,{\rm fs})^{-1}\approx 265\,{\rm cm}^{-1}$.

\begin{table}[b]
\caption{${\rm Q}_y$-transition dipole moments of the FMO complex of {\it C. tepidum}.}
\label{Table_FMO}
\begin{ruledtabular}
\begin{tabular}{lllllllll}
Site & 1 & 2 & 3 & 4 & 5 & 6 & 7 \\
\hline
x-component & 0.741 & 0.857 & 0.197 & 0.799 & 0.736 & 0.135 & 0.495 \\
y-component & 0.560 & -0.503 & -0.957 & 0.533 & -0.655 & 0.879 & 0.708 \\
z-component & 0.369 & 0.107 & 0.210 & 0.276 & -0.164 & -0.456 & 0.503
\end{tabular}
\end{ruledtabular}
\end{table}

\section{Theory of absorption spectra}

Within the Franck-Condon approximation, the linear absorption spectrum of isotropic samples is described by~\cite{SI_Mukamel1995}
\begin{equation}
    A(\omega)\propto  {\rm Re}{\int_{0}^{\infty}} dt e^{i\omega t} d(t)={\rm Re}{\int_{0}^{\infty}} dt e^{i\omega t}{\rm tr}[{\bm \mu}\cdot (e^{-iHt}{\bm \mu}\rho(0)e^{iHt})],
    \label{eq:absorption}
\end{equation}
with $d(t)$ denoting dipole-dipole correlation function and $\rho(0)=\ket{g}\bra{g}\otimes\rho_{v}(T)$ an initial state where $\ket{g}$ is the global electronic ground state and $\rho_{v}(T)$ the thermal state of vibrational modes at temperature $T$~\cite{SI_Mukamel1995}. The transition dipole moment operator ${\bm \mu}$ is modelled by
\begin{equation}
    {\bm \mu}=\sum_{i=1}^{N}{\bm \mu}_{i}(\ket{\epsilon_i}\bra{g}+\ket{g}\bra{\epsilon_i}),
\end{equation}
with ${\bm \mu}_{i}$ representing the transition dipole moment of site $i$. In the FMO simulations, the transition dipole moments of sites 1-7 summarized in Table~\ref{Table_FMO} were considered. In dimer simulations, the transition dipole moments of two pigments were assumed to be mutually orthogonal (${\bm \mu}_1 \cdot {\bm \mu}_2=0$), so that both low- and high-energy excitons are bright states. The lifetime of optical coherences, ${\rm tr}_{v}[e^{-iHt}{\bm \mu}\rho(0)e^{iHt}]$, and corresponding absorption line widths are determined by the homogeneous and inhomogeneous broadenings induced by the vibronic coupling $H_{e-v}$ and static disorder, respectively.

\section{Simulation-time dependence of coarse-graining}

In the main manuscript, we considered the bath correlation function (BCF) of the full FMO complex up to $300\,{\rm fs}$ to construct an effective spectral density. This is the time scale of absorption measurements, namely the lifetime of optical coherences between electronic ground and excited states. It is found that 62 intra-pigment modes present in the actual FMO spectral density can be effectively described by five Lorentzian peaks using our course-graining scheme, which can be estimated based on the number of peaks present in the frequency spectrum of the BCF up to $300\,{\rm fs}$, as shown in Fig.~\ref{Fig1}(a).

As the time scale of interest increases, a larger number of peaks appears in the frequency spectrum of the BCF. In two-dimensional (2D) electronic spectroscopy, a molecular sample is perturbed by three femtosecond laser pulses with controlled time delays~\cite{SI_JonasARPC2003,SI_Brixner2004,SI_LimPRL2019}. The time delay $t_1$ between first and second pulses enables one to monitor the phase evolution of the optical coherences created by the first pulse, decaying within $\sim 300\,{\rm fs}$ as is the case of linear absorption. The molecular dynamics, such as energy/charge transfer and vibrational motion, is monitored as a function of the time delay $t_2$ between second and third pulses. In typical 2D experiments on PPCs, $t_2 \lesssim 1\,{\rm ps}$ has been considered to study photosynthetic energy/charge transfer and quantum coherence effects occurring on a sub-ps time scale~\cite{PanitchayangkoonPNAS2010,ColliniN2010,RomeroNP2014}. The optical coherences created by the third laser pulse induce oscillating transition dipole moments generating nonlinear 2D signals. The optical coherences decay within $\sim 300\,{\rm fs}$, described by an additional time variable $t_3$. In simulations, one needs to compute the time evolution of a molecular system as a function of time $t=t_1+t_2+t_3 \lesssim 2\,{\rm ps}$, leading to 2D electronic spectra when the time-domain simulated data are Fourier-transformed with respect to $t_1$ and $t_3$. Therefore, one needs to simulate the dynamics of a molecular system on a picosecond time scale, namely $\sim 2\,{\rm ps}$, to compute 2D electronic spectra.

In Fig.~\ref{Fig1}(b) and (c), respectively, we consider the typical time scale of the energy transfer in PPCs ($1\,{\rm ps}$) and the time scale associated with the simulations of 2D electronic spectroscopy ($2\,{\rm ps}$). It is notable that the number of peaks present in the frequency spectrum is increased from 5, via 16, to 28, as the time scale of interest increases from $300\,{\rm fs}$, via $1\,{\rm ps}$, to $2\,{\rm ps}$, but they are still smaller than the total number of intra-pigment modes in the full FMO spectral density ($M=62$). Therefore, our coarse-graining scheme becomes less efficient as the time scale of interest increases, but it still requires less computational resources than the full FMO spectral density.

\begin{figure}
\includegraphics[width=0.8\textwidth]{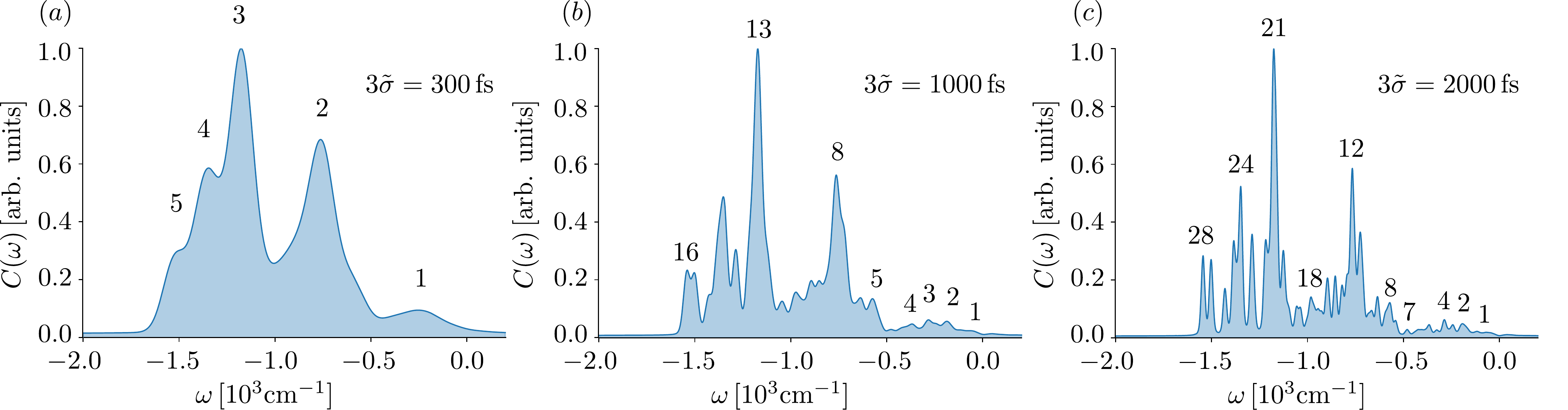}
\caption{(a)-(c) Fourier transform (FT) of the bath correlation function $C(t)$ of the full FMO spectral density at $77\,{\rm K}$ weighted by a Gaussian function $e^{-t^{2}/2\tilde{\sigma}^2}$ with $3\tilde{\sigma}\in\{300\,{\rm fs}, 1\,{\rm ps}, 2\,{\rm ps}\}$. The number of peaks is increased from 5, via 16, to 28 as the time scale of interest is increased from $300\,{\rm fs}$ (absorption measurements), via $1\,{\rm ps}$ (energy transfer dynamics), to $2\,{\rm ps}$ (nonlinear spectroscopy measurements).}
\label{Fig1}
\end{figure}

\begin{figure}
\includegraphics[width=0.8\textwidth]{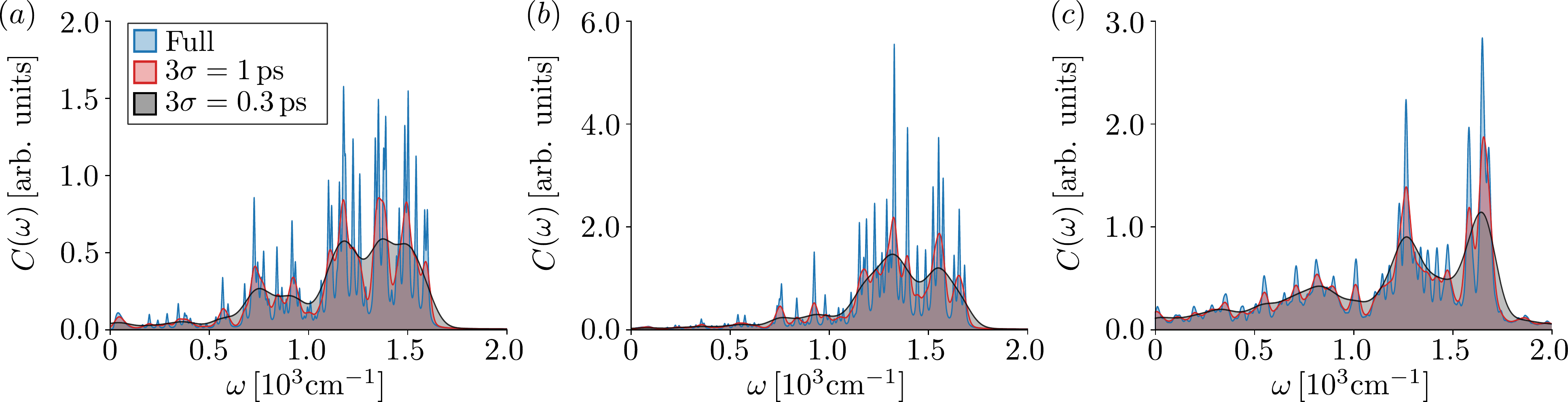}
\caption{Fourier transform (FT) of the bath correlation function $C(t)$ at $77\,{\rm K}$ of (a) the special pair in bacterial reaction centre, (b) WSCP from cauliflowers, and (c) PCB158c pigment in the PC645 complex from marine algae. The FT of the BCF without a Gaussian filter is shown in blue, while the FT of the BCF weighted by a Gaussian function $e^{-t^{2}/2\tilde{\sigma}^2}$ with $3\tilde{\sigma} = 300\,{\rm fs}$ ($3\tilde{\sigma} = 1\,{\rm ps}$) is shown in black (red).}
\label{Fig2}
\end{figure}

So far, we have shown that the complexity of an effective environment, capturing the full environmental effects induced by the experimentally determined phonon spectral density of the FMO complex, can be significantly reduced by considering a finite time scale. Here, we demonstrate that these observations are also valid for other photosynthetic pigment-protein complexes. In Fig.~\ref{Fig2}(a)-(c), respectively, we consider experimentally determined phonon spectral density of the special pair (SP) in bacterial reaction centre~\cite{SI_ZazubovichJPCB2001}, that of water-soluble chlorophyll-binding protein (WSCP) from cauliflower~\cite{SI_PieperJPCB2011}, and theoretically computed phonon spectral density of the PCB158c pigment present in the PC645 complex from marine algae~\cite{SI_BlauPNAS2018}. These complexes have all highly structured phonon spectral densities, as shown in blue, where the multi-peak structure originates from the discrete vibrational spectrum of the intra-pigment modes~\cite{SI_StuartWiley2005}. For all the three cases, the effective environments constructed based on a finite time scale $300\,{\rm fs}$ ($1\,{\rm ps}$) consist of $\sim 8$ ($\sim 15$) peaks, as shown in black (red), similar to the case of the FMO complex. These results demonstrate that the computational costs of non-perturbative simulations of PPCs can be reduced by our coarse-graining scheme, which is not limited to the FMO complex considered in the main manuscript. We note that our coarse-graining scheme does not depend on the electronic energy-level structures of PPCs, or more generally on the parameters of open-quantum systems. Rather, it solely depends on the BCF and finite time scale of interest.

\section{Time scale analysis of absorption spectra of the FMO complex}

In the main manuscript, we constructed an effective environment based on the FT of the BCF weighted by a Gaussian function $e^{-t^2/2 \tilde{\sigma}^2}$ where the standard deviation $\tilde{\sigma}$ is taken to be $3\tilde{\sigma} = 300\,{\rm fs}$, so that the Gaussian-weighted BCF decays within $300\,{\rm fs}$. This is based on the assumption that ensemble-averaged optical coherences decay within $300\,{\rm fs}$, which can be easily estimated based on the simulations in the absence of environments where the dephasing of optical coherences is induced solely by static disorder. In case the time scale $\tau$ of the optical coherence dynamics determining absorption lineshapes is shorter than $300\,{\rm fs}$, it is sufficient to construct an effective environment with a reduced standard deviation $3\tilde{\sigma} = \tau < 300\,{\rm fs}$ for numerically exact simulations of absorption spectra.

\begin{figure}
\includegraphics[width=0.8\textwidth]{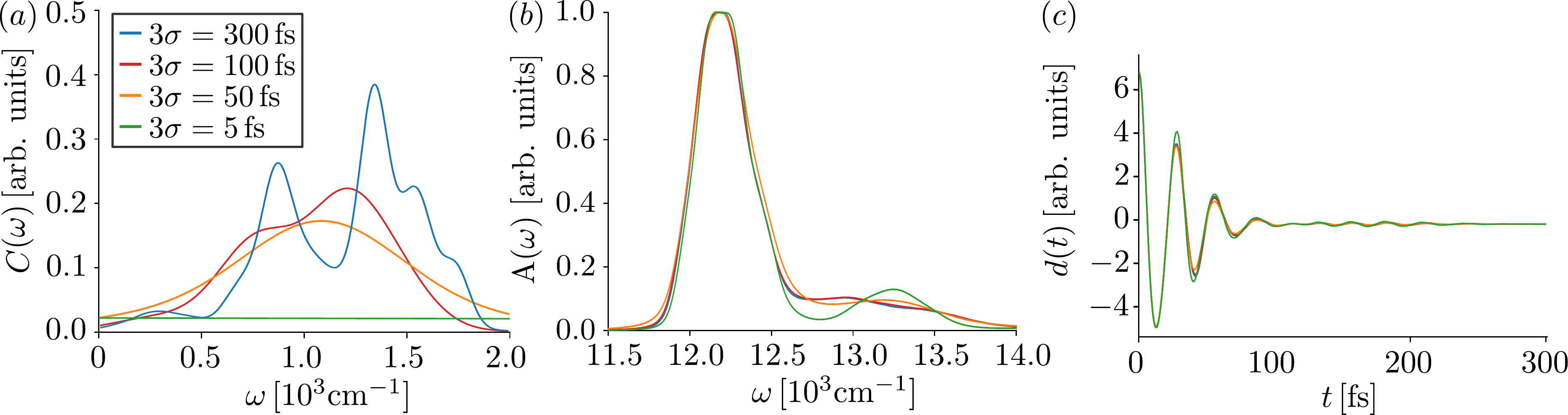}
\caption{(a) Fourier transform of the Gaussian-filtered bath correlation function of the full FMO spectral density at $77\,{\rm K}$ where the standard deviation of the Gaussian function is taken to be $3\tilde{\sigma} = 300\,{\rm fs}$ (blue), $100\,{\rm fs}$ (red), $50\,{\rm fs}$ (orange) and $5\,{\rm fs}$ (green). (b) Absorption spectra of the FMO complex and (c) corresponding dipole-dipole correlation functions $d(t)$ obtained using the four effective environments shown in (a).}
\label{Fig3}
\end{figure}

To investigate how small standard deviations can reproduce the numerically exact absorption spectra of the FMO complex at $77\,{\rm K}$, we constructed effective environments for $3\tilde{\sigma}\in\{5,50,100,300\}\,{\rm fs}$. The results are shown in Fig.~\ref{Fig3}(a). As the standard deviation is decreased from $3\tilde{\sigma}=300\,{\rm fs}$ to $3\tilde{\sigma}=5\,{\rm fs}$, the structure of the effective environments becomes simpler and broader. As shown in Fig.~\ref{Fig3}(b), the numerically exact absorption spectra can be well reproduced when $3\tilde{\sigma}\gtrsim 100 \,{\rm fs}$. The numerical results obtained using $3\tilde{\sigma}\ll 100 \,{\rm fs}$ deviate from numerically exact ones, particularly in the high-energy domain above $12.5\times 10^3 {\rm cm}^{-1}$. These results can be rationalized by analysing optical coherence dynamics. In Fig.~\ref{Fig3}(c), the dipole-dipole correlation function $d(t)$ in the presence of static disorder is shown as a function of time $t$. Note that $d(t)$ mainly decays within $\sim 100\,{\rm fs}$, with small fluctuations persisting up to $\sim 300\,{\rm fs}$. Since absorption spectrum is proportional to the Fourier transform of $d(t)$, the large-amplitude dynamics of $d(t)$ at early times $t\lesssim 100\,{\rm fs}$ may play a crucial role in determining the absorption spectra. This is in line with the numerical results shown in Fig.~\ref{Fig3}(b) where numerically exact absorption spectra are well described by effective environments constructed using $3\tilde{\sigma}\gtrsim 100 \,{\rm fs}$. These results demonstrate that the performance of our coarse-graining scheme can be further enhanced by optimizing the time scale $\tau=3\tilde{\sigma}$ of open-system dynamics governing experimentally observable quantities, although this necessitates repetitions of non-perturbative simulations.

\begin{figure}
\includegraphics[width=0.8\textwidth]{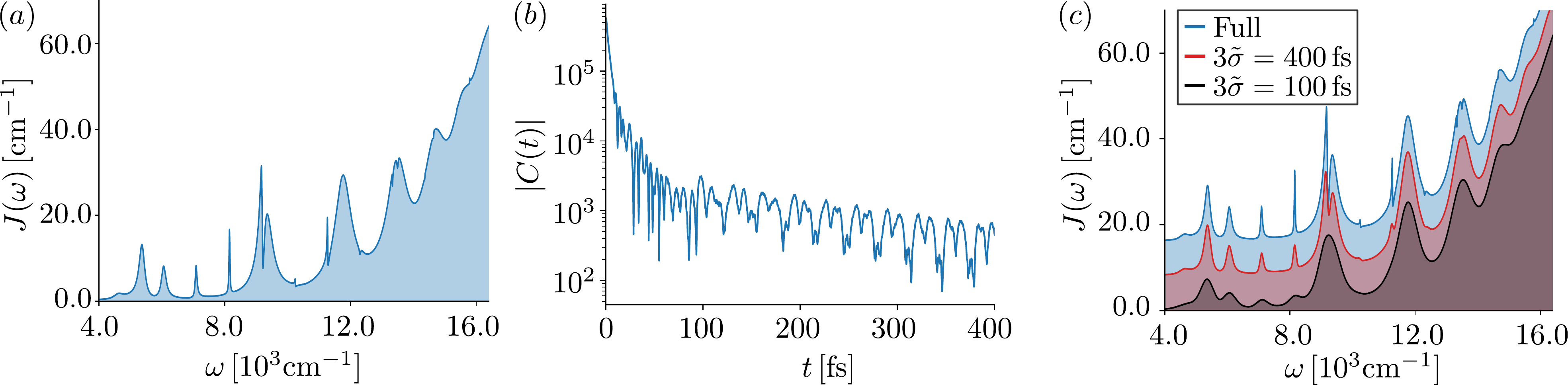}
\caption{(a) Photon spectral density of the electromagnetic environment of a silver dimer nanoantenna embedded in a dielectric microsphere~\cite{SI_MedinaPRL2021}. (b) Corresponding bath correlation function (BCF) at zero temperature. (c) Fourier transform of the full BCF (blue) and those of Gaussian-filtered BCFs with $3\tilde{\sigma} = 400\,{\rm fs}$ (red) or $3\tilde{\sigma} = 100\,{\rm fs}$ (black). The offset values of the FTs are shifted for better visibility.}
\label{Fig4}
\end{figure}

\newpage

\section{Coarse-graining of emitter-photon interaction}

In this work, we demonstrated that the structure of phonon spectral densities of photosynthetic pigment-protein complexes, or molecular systems in general, can be systematically coarse-grained while preserving full environmental effects on a timescale of interest. Our coarse-graining scheme can be applied to any harmonic environments, which are initially in a thermal state and linearly coupled to open systems. A relevant example is provided by polaritonic systems where an emitter is coupled to photonic environments. As shown in Ref.~\cite{SI_MedinaPRL2021}, the emitter-photon coupling spectrum of polaritonic systems is characterized by a high degree of structure, reproduced in Fig.~\ref{Fig4}(a). In Ref.~\cite{SI_MedinaPRL2021}, the emitter dynamics up to $400\,{\rm fs}$ was simulated using a non-perturbative method where the continuous photonic environment is described by a finite number of discrete modes under Lindblad noise. Due to the highly structured emitter-photon coupling spectrum, characterised by sharp Fano-like profiles, the interaction between discrete modes was introduced to take into account environmental effects with high accuracy, albeit introducing increased computational costs~\cite{SI_MedinaPRL2021}. As shown in Fig.~\ref{Fig4}(b), the bath correlation function of the photonic environment mainly decays within $100\,{\rm fs}$, due to the broad frequency spectrum of the photon spectral density shown in Fig.~\ref{Fig4}(a), with small amplitudes persisting up to $400\,{\rm fs}$. This implies that the emitter dynamics is mainly governed by the bath correlation function at early times (below $100\,{\rm fs}$). In Fig.~\ref{Fig4}(b), the effective spectral densities obtained by our coarse-graining scheme with $3\tilde{\sigma}\in\{100,400\}\,{\rm fs}$ are shown. Note that the narrow Fano-like features are significantly suppressed by our scheme, which may enable one to perform numerically exact simulations without resorting to consider the interaction between discrete modes, thus reducing computational costs.

\vspace{1cm}

\section{Computational costs of DAMPF: the FMO complex and linear chains}

\begin{figure}
\includegraphics[width=0.8\textwidth]{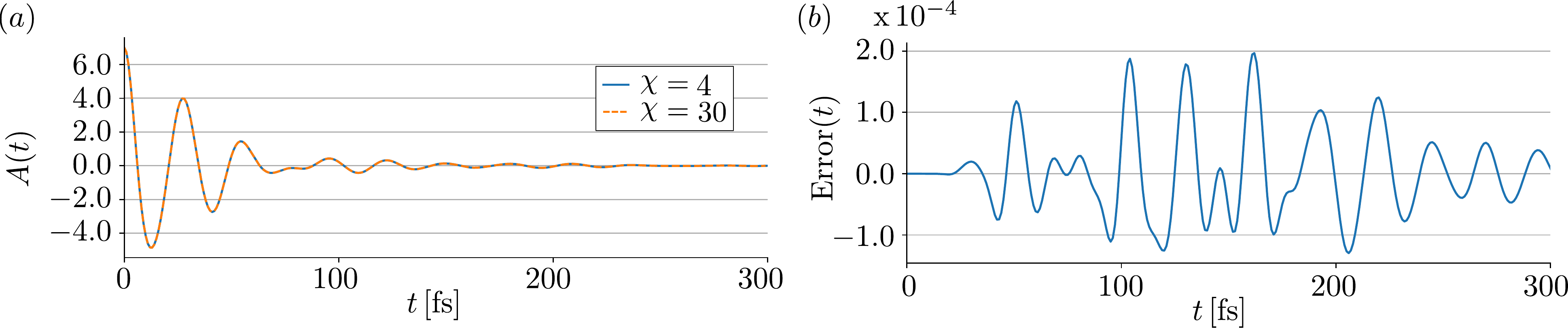}
\caption{(a) Optical coherence dynamics of the full FMO model ($N=7$ sites and $M=62$ intra-pigment modes per site) weighted by a Gaussian function $e^{-t^2/2\tilde{\sigma}^2}$ with $3\tilde{\sigma}=300\,{\rm fs}$, computed by DAMPF with bond dimensions $\chi=4$ and $\chi=30$. (b) The difference between DAMPF results obtained with $\chi=4$ and $\chi=30$.}
\label{Fig5}
\end{figure}

In this section we discuss the computational costs of DAMPF for absorption simulations of the FMO complex and linear chains.

\begin{figure}
\includegraphics[width=0.5\textwidth]{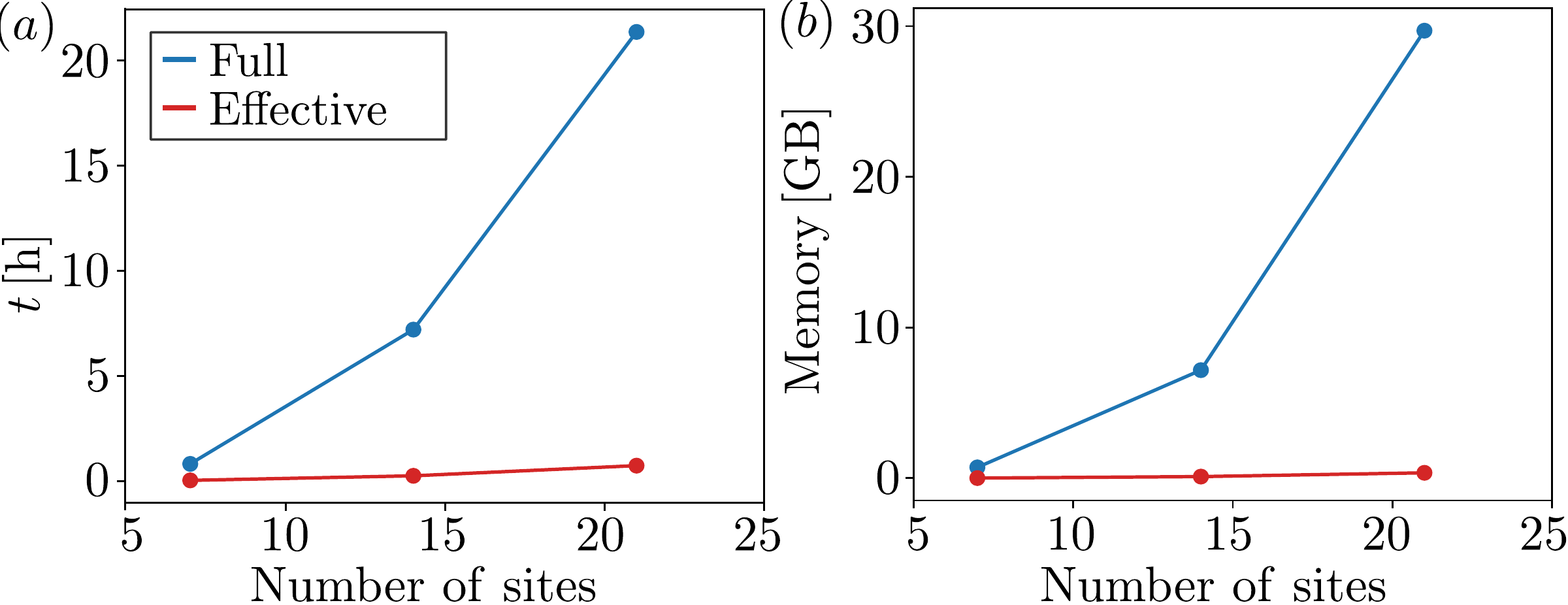}
\caption{(a) Computational time and (b) memory required for DAMPF simulations of the absorption spectra of the linear chains consisting of 7, 14, 21 sites (see the main text for more details).}
\label{Fig6}
\end{figure}

\begin{table}[b]
\caption{Site energies of a seven-site linear chain considered in Fig.~\ref{Fig6}.}
\label{Table_chain}
\begin{ruledtabular}
\begin{tabular}{llllllll}
Site & 1 & 2 & 3 & 4 & 5 & 6 & 7 \\
\hline
$\epsilon_i\,[{\rm cm}^{-1}]$ & 1141 & 1478 & 1031 & 1233 & 1163 & 1420 & 1282
\end{tabular}
\end{ruledtabular}
\end{table}

In Fig.~\ref{Fig5}, we show the optical coherence dynamics of the full FMO model, consisting of $N=7$ pigments and $M=62$ intra-pigment modes per site, weighted by a Gaussian function
\begin{equation*}
    A(t) = {\rm tr}[{\bm \mu}\cdot (e^{-iHt}{\bm \mu}\rho(0)e^{iHt})] e^{-t^2/2\tilde{\sigma}^2}
\end{equation*}
with $3\tilde{\sigma}=300\,{\rm fs}$. In simulations, the mean site energies of the FMO complex~\cite{SI_RengerBJ2006} were considered. Here the Gaussian function makes $A(t)$ decay within $300\,{\rm fs}$, so that the Fourier transform of $A(t)$ leads to well-defined absorption line shapes without ringing artifacts. As shown in Fig.~\ref{Fig5}, $A(t)$ computed up to $300\,{\rm fs}$ by DAMPF with bond dimensions $\chi=4$ and $\chi=30$ are quantitatively well-matched, implying that low bond dimensions $\sim 4$ are sufficient to obtain numerically converged absorption spectra of the full FMO model. For the effective environment considered in the main manuscript, we found that the bond dimension required for numerically exact DAMPF simulations of the absorption spectra of the FMO complex is $\chi=2$.

In Fig.~\ref{Fig6}, we considered linear molecular chains consisting of 7, 14, 21 sites. For simplicity, we assumed that the electronic coupling $V=50\,{\rm cm}^{-1}$ between nearby sites is uniform and each site is coupled to a local phonon environment at $77\,{\rm K}$ described by the experimentally determined phonon spectral density of the FMO complex. For the seven-site chain, we considered randomly generated site energies with detunings being smaller in magnitude than $500\,{\rm cm}^{-1}$, as summarised in Table~\ref{Table_chain}. For the longer chains, we repeatedly used the site energies in Table~\ref{Table_chain}, treating them as if multiple seven-site chains were connected. In Fig.~\ref{Fig6}(a) and (b), respectively, the computational time and memory required for DAMPF simulations of the absorption spectra of the linear chains are shown. Here the computational costs for dealing with the full FMO spectral density and effective environments, respectively, are shown in blue and red. For the linear chains consisting of 7, 14, 23 sites, respectively, the computational costs are reduced from 51, 434, 1283 min to 4, 17, 46 min (from 0.7, 7, 29 GB to 7, 105, 355 MB) when the effective environment is considered instead of the full FMO spectral density (all simulations were executed using 15 cores in an Intel Xeon 6252 Gold CPU). The bond dimensions required for numerically exact DAMPF simulations of the linear chains consisting of 7, 14, 21 sites are 3, 4, 4 (9, 10, 11), respectively, when the effective (full) environment is considered, which is larger than the FMO case. These results demonstrate that our coarse-graining scheme combined with DAMPF enables one to perform numerically exact simulations of large vibronic systems consisting of several tens of sites where each site is coupled to a highly structured phonon environment at finite temperatures.

\end{widetext}